\documentclass[a4paper]{article}
\usepackage{spconf,amsmath,epsfig}
\usepackage[skip=1pt]{caption}
\usepackage{bm}
\usepackage{makecell}
\usepackage{cite}
\newcommand\numberthis{\addtocounter{equation}{1}\tag{\theequation}}

\title{IMAGE DENOISING WITH LESS ARTEFACTS: \\
NOVEL NON-LINEAR FILTERING ON FAST PATCH REORDERINGS}
\name{Kireeti Bodduna and Joachim Weickert
\thanks{J.W. has received funding from the European 
Research Council (ERC) under the European Union's Horizon 
2020 programme (grant agreement 
no.~741215, ERC Advanced Grant INCOVID).
We thank Utz Ermel, Lasse Sprankel and Prof. Achilleas Frangakis 
from the Electron Microscopy Group at the 
University of Frankfurt for providing us with data.
We also thank our colleague Tobias Alt  
for useful comments on a draft version of the paper. }}
\address{Mathematical Image Analysis Group, \\
	Saarland University, 66041 Saarbr{\"u}cken, Germany.\\
\{bodduna,weickert\}@mia.uni-saarland.de}
\begin{document}
%
\maketitle
\begin{abstract}
Leading denoising methods such as 3D block matching (BM3D)
are patch-based. However, they can suffer from frequency domain 
artefacts and require to specify explicit noise models.
We present a patch-based method that avoids 
these drawbacks. It combines a simple 
and fast patch reordering with a non-linear smoothing. 
The smoothing rewards both patch and 
pixel similarities in a multiplicative way.
We perform experiments on real world images with 
additive white Gaussian noise (AWGN), and on 
electron microscopy data with a more general additive noise 
model. Our filter outperforms BM3D in 77\% of the experiments, 
with improvements of up to 29\%  with respect to the mean 
squared error. 
\end{abstract}
\begin{keywords}
non-local patch-based methods, 
diffusion methods, image denoising, additive white 
Gaussian noise
\end{keywords}
\section{Introduction}
\label{sec:intro}
Non-local patch-based methods \cite{BCM05,LBM2013,
DFKE07, ram2013image, ram2013imageNL, pierazzo2015da3d} 
have been producing superior image denoising results 
since quite a few years now. These models offer two main 
advantages: Firstly, the assumption 
that similar pixels have similar neighbourhoods 
around them, is quite robust in a noisy scenario. 
Secondly, one can access similar data from distant 
regions in the image. 

The non-local Bayes (NLB) \cite{LBM2013a, LBM2013}
and BM3D \cite{DFKE07, Lebrun2012}
approaches produce state-of-the-art results. 
NLB uses Bayesian modelling while BM3D 
employs Fourier domain filtering. However, 
it is well known that when the assumptions about noise-free 
images are violated, one can see artefacts in the denoised 
images. The idea that we can find completely similar patches 
within an image in general need not be satisfied. 
Thus, there is a risk that data from dissimilar regions 
can diffuse into each other, which leads to the 
above mentioned artefacts. 
This observation is well documented for both 
NLB and BM3D \cite{pierazzo2015da3d}. 

One remedy to eliminate these artefacts 
is to use an additional post-processing step 
\cite{pierazzo2015da3d}. 
Another possibility which was proposed by Ram et al. 
\cite{ram2013image, ram2013imageNL}
is to employ a smooth reordering of patches 
for subsequent filtering. However, the underlying reason why 
the latter method is better than BM3D is not well understood. 
It appears plausible to us that it minimises information 
exchange between dissimilar regions with the help of patch 
reordering, thus reducing the artefacts and consequently 
leading to better results. However, this comes at a 
computationally very expensive reordering step, which 
basically requires to solve a travelling salesman problem. 

{\bf Our Contribution.} We introduce a 
new method to solve the above artifact problem 
without the need of an additional post-processing step 
and also within a relatively low computational time.
In contrast to the methods of Ram et al. \cite{ram2013image, 
ram2013imageNL}, we use a simpler and much faster 
patch reordering, and combine it with a more sophisticated 
non-linear filtering. Hence, we call our method 
\textit{non-linear filtering on fast patch 
reorderings (NFPR)}. In particular, we employ a filtering technique 
which is a novel combination of weights that 
reward both patch and pixel similarities. 
Moreover, we always use disc-shaped 
windows, thus leading to a 
rotationally invariant model. In contrast to NLB and BM3D, 
we avoid an explicit AWGN assumption and hence 
are more robust with respect to the noise type.

{\bf Paper Structure.} In Section \ref{sec:modelling}, we 
introduce our proposed NFPR framework for noise elimination 
along with proper motivations. 
In Section \ref{sec:expAndDisc}, we showcase 
a comparative evaluation of NFPR with NLB, BM3D and 
the method of Ram et al. \cite{ram2013image}, for both
real-world test images and electron microscopy data. 
Finally, in Section \ref{sec:concAndOutlook}, 
we conclude with a summary of our contribution and 
an outlook to future work. 


\begin{table}[t]
\setlength{\tabcolsep}{3pt}
\small
\centering
\begin{tabular}{ l c  c c  r r r}
 \hline 
 Image & $\sigma$ & $\lambda$ 
 & $k_{\textrm{max}}$ & NFPR & NLB & BM3D \\
 \hline 
 L40 & 150 & 11.5 & 16 & 74.00 & 69.30 & \textbf{68.27} \\  
 L60 & 160 & 15.5 & 16 & \textbf{104.58} & 109.3 & 104.83\\  
 L80 & 175 & 20.0 & 14 & \textbf{139.37} & 154.40 & 143.23\\ 
 L100 & 175 & 23.5 & 16 & \textbf{164.85} & 198.89 & 183.78 \\ 
 L120 & 190 & 27.0 & 15 & \textbf{196.96} & 254.90 & 228.39 \\  
  \vspace{0.5em}
 L140 & 195 & 31.5 & 15 & \textbf{231.48} & 312.95 & 273.09 \\  
 B40 & 130 & 15.0 & 13 & 254.87 & 233.36 & \textbf{233.34}  \\  
 B60 & 160 & 20.5 & 8 & 333.37 & 333.22 & \textbf{315.37} \\  
 B80 & 165 & 26.0 & 9 & 400.71 & 425.61 & \textbf{391.02} \\  
 B100 & 180 & 30.5 & 9 & 457.70 & 486.65 & \textbf{453.37} \\  
 B120 & 180 & 34.5 & 10 & \textbf{505.86} & 551.94 & 512.38 \\  
               \vspace{0.5em}
 B140 & 190 & 36.5 & 11 & \textbf{556.54}& 616.51 & 572.44 \\   
 H40 & 140 & 10.5 & 27 & 58.77 & 62.16 & \textbf{56.01}\\ 
 H60 & 160 & 12.0 & 27 & \textbf{81.24} & 104.85 & 92.15 \\ 
 H80 &  180 & 15.0 & 23 & \textbf{116.63} & 156.88 & 130.64 \\ 
 H100 & 185 & 17.5 & 17 & \textbf{153.27} & 218.91 & 187.60 \\
 H120 & 205 & 22.5 & 17 & \textbf{192.11} & 289.12 & 235.53 \\  
  \vspace{0.5em} 
 H140 & 200 & 25.0 & 19 & \textbf{233.62} & 356.59 & 300.88 \\    
 P40 & 155 & 11.5 & 16 & \textbf{57.73} & 60.03 & 58.91 \\ 
 P60 & 160 & 16.0 & 17 & \textbf{83.73} & 95.22 & 91.16 \\ 
 P80 & 185 & 18.5 & 15 & \textbf{112.39} & 128.95 & 124.08 \\ 
 P100 & 195 & 21.0 & 15 & \textbf{139.05} & 171.88 & 160.88 \\ 
 P120 & 205 & 25.0 & 15 & \textbf{167.70} & 216.02 & 200.58 \\ 
 P140 & 205 & 29.5 & 15 & \textbf{198.83} & 266.79 & 241.20 \\    
 \hline 
 \end{tabular}
 \vspace{1em}
 \captionof{table}{MSE values of denoised images including 
 NFPR parameters. 
 L40 stands for Lena image with 
 $\sigma_{\textrm{noise}}$ = 40. 
 B, H, P denote Bridge, House and Peppers respectively.}
 \label{table1}
\end{table}

\section{Modelling of our denoising algorithm}
\label{sec:modelling}

Our NFPR technique consists of two parts: 
The goal of the first step is to achieve a 
fast patch-based reordering of the pixels.
In the second step, we employ a non-linear smoothing on the 
reordered pixels which yields the denoised image. 
In the following we describe these steps in detail.

\noindent\textbf{Step 1~:~Fast Patch Reordering.} 
In order to compute a smooth reordering of pixels, 
we employ a patch-based similarity approach: We 
first consider a disc-shaped search region $B_\textrm{search}$ 
of radius $\rho_\textrm{search}$
around every pixel $u_i$ in the 2D image domain. 
We then compute the $L_2$ norm 
$d_{ij}$ between disc-shaped patches of radius $\rho_{\textrm{sim}}$,
centered around $u_i$ and $u_j,$ for all $j 
\in B_\textrm{search}$. 
This is followed by constructing a set $P_i$ of $N$ pixels 
within $B_\textrm{search}$ which have the least distance from 
$u_i$ according to $d_{ij}$. 
This set characterises 
the desired smooth reordering of pixels. 
In contrast to Ram et al. \cite{ram2013image, 
ram2013imageNL} who solve instances of the NP-hard 
travelling salesman problem, we compute the reordering using 
just a simple sort operation. 
Ideally, when we average noisy versions of the 
same greyvalue, we should not introduce artefacts. 
However, we have noisy versions of approximately equal grey values 
in the set $P_i$. Moreover, the above simple reordering
is achieved at the cost of some disordered pixels in $P_i$, 
that come from areas of dissimilar greyvalues. In the second 
step of the algorithm, we employ a very robust non-linear smoothing 
technique, to deal with both problems. \\

\noindent\textbf{Step 2 : Non-linear Smoothing.} 
The goal of this step is to optimally combine the set of pixels 
$P_i$, obtained from the first step, 
and compute a final denoised image. 
To this end, we apply a non-linear smoothing process on this
set. This can be thought of as diffusing information
between pixels in a space defined by the neighbourhood 
similarity distances $d_{ij}$ instead of the generally used spatial 
distances. We utilise two assumptions which form the core of 
our structure preserving smoothing technique:   
Firstly, similar pixels have relatively 
smaller absolute tonal differences 
$|u_j - u_i|$. 
Secondly, they also have similar neighbourhoods around them. 
Although we have already used such an idea for patch reordering, 
we will be re-using the distances $d_{ij}$ through 
a multiplicative combination of both assumptions. 
This gives us an advantage in scenarios where 
one of the assumptions might be violated in the presence of
noise. The discrete evolution equation for 
our smoothing process is given by
\begin{align*}
\label{discretization}
&\frac{u_i^{k+1} - u_i^{k}}{\tau} =  
a_i^k \cdot \left(
\sum_{\substack{j \in P_i^k} }
g \left({u^{k}_{\sigma j}} - 
{u^{k}_{\sigma i}}\right) 
h \left( d_{ij}^k \right)
\left( u_j^k - u_i^k \right) \right.  \\ 
&+ \left.  \sum_{\substack{j \in P^{\textrm{add},k}_{i}} }
g \left({u^{k}_{\sigma j}} - 
{u^{k}_{\sigma i}}\right) 
 h \left( d_{ij}^k \right)
\left( u_j^k - u_i^k \right) \right) \numberthis.
\end{align*}
This equation has two terms on the right hand side, 
which model two types of information exchange: 
Remember that $P_i$ denotes the set of pixels which are closest 
to pixel $u_i$ according to the distance $d_{ij}$. 
Every pixel in the image will have its own reordered set. Thus, 
the pixel $u_i$ could also be part of sets other than $P_i$. 
The symbol $P^{\textrm{add}}_{i}$ denotes an additional 
set of pixels in whose corresponding reordered sets, 
$u_i$ is present. 
The two terms mentioned in the above equation represent 
interactions with these two sets of pixels $P_i$ and
$P^{\textrm{add}}_{i}$, respectively.
This can also be seen as collaborative filtering similar to 
BM3D and NLB. 

\begin{table}[t]
\setlength{\tabcolsep}{3pt}
\small
\centering
\begin{tabular}{ l c  c c  r r }
 \hline 
 Image & $\sigma$ & $\lambda$ 
 & $k_{\textrm{max}}$ & NFPR & REC   \\
 \hline 
 L50 & 150 & 14.5 & 17 & 91.36 & \textbf{82.90} \\  
 L75 & 170 & 19.0 & 15 & 126.90 & \textbf{123.75} \\  
  \vspace{0.5em}
 L100 & 175 & 23.5 & 16 & 164.85 & \textbf{163.05}  \\   
 H50 & 160 & 11.0 & 23 & \textbf{70.94} & 74.45 \\ 
 H75 & 165 & 15.5 & 23 & \textbf{108.71} & 120.41 \\ 
 H100 & 185 & 17.5 & 17 & \textbf{153.27} & 167.52 \\   
 \hline 
 \end{tabular}
 \vspace{1em}
 \captionof{table}{MSE values of denoised images including 
 NFPR parameters. Abbreviations as in Table \ref{table1}, 
 REC - Ram et al. \cite{ram2013image}.}
 \label{table2}
\end{table}
%
\begin{figure*}[t]
  \large{ \hspace{11mm} noisy  \hspace{20mm}  
  NLB   \hspace{20mm}  BM3D 
  \hspace{20mm}  NFPR  \hspace{18mm} original  } \\
  \centering
  \includegraphics[width=0.17\linewidth]
  {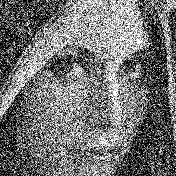}\hspace{0.2em}
  \includegraphics[width=0.17\linewidth]
  {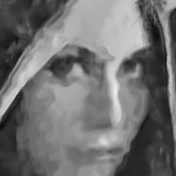}\hspace{0.2em}
  \includegraphics[width=0.17\linewidth]
  {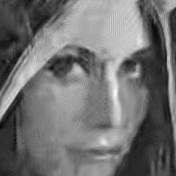}\hspace{0.2em}
  \vspace{0.3em}
  \includegraphics[width=0.17\linewidth]
  {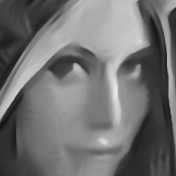}\hspace{0.2em}
  \includegraphics[width=0.17\linewidth]
  {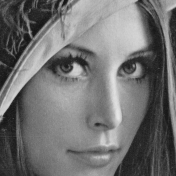}
  \includegraphics[width=0.17\linewidth]
  {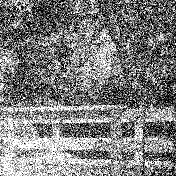}\hspace{0.2em}
    \includegraphics[width=0.17\linewidth]
  {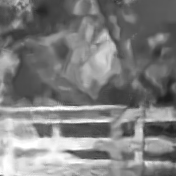}\hspace{0.2em}
  \includegraphics[width=0.17\linewidth]
  {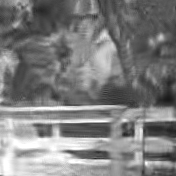}\hspace{0.2em}
    \vspace{0.3em}
  \includegraphics[width=0.17\linewidth]
  	{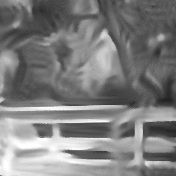}\hspace{0.2em}
  \includegraphics[width=0.17\linewidth]
  {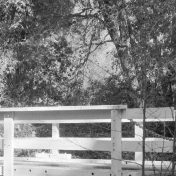}  					
  \includegraphics[width=0.17\linewidth]
  {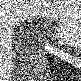}\hspace{0.2em}
    \includegraphics[width=0.17\linewidth]
  {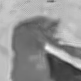}\hspace{0.2em}
  \includegraphics[width=0.17\linewidth]
  {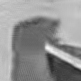}\hspace{0.2em}
    \vspace{0.3em}
  \includegraphics[width=0.17\linewidth]
  {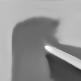}\hspace{0.2em}  
  \includegraphics[width=0.17\linewidth]
  {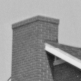}
  \includegraphics[width=0.17\linewidth]
  {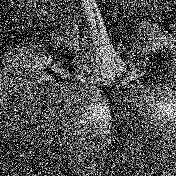}\hspace{0.2em}
  \includegraphics[width=0.17\linewidth]
  {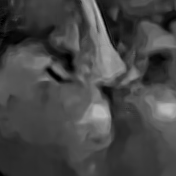}\hspace{0.2em}
  \includegraphics[width=0.17\linewidth]
  {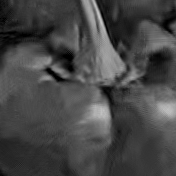}\hspace{0.2em}
  \vspace{0.3em}
  \includegraphics[width=0.17\linewidth]
  {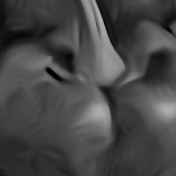}\hspace{0.2em}    					
  \includegraphics[width=0.17\linewidth]
  {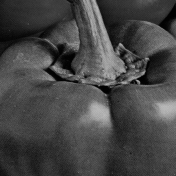}
  					
\caption{\textbf{Top to Bottom:} Zoom into  
Lena, Bridge, House and Peppers images ($\sigma_{\mathrm{noise}} = 80$).}
\label{fig:res1}
\end{figure*}

\begin{figure*}[t]
  \large{ \hspace{0.1mm} noisy  \hspace{32mm}  
  NFPR \hspace{53mm} FRC plot} \\
  \centering
  \hspace{3em}\includegraphics[width=0.24\linewidth]
  {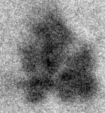}\hspace{0.2em}
  \includegraphics[width=0.24\linewidth]
  {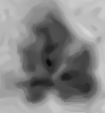} \hspace{2em}
    \vspace{0.3em}
  	\includegraphics[width=0.4\linewidth]
  	{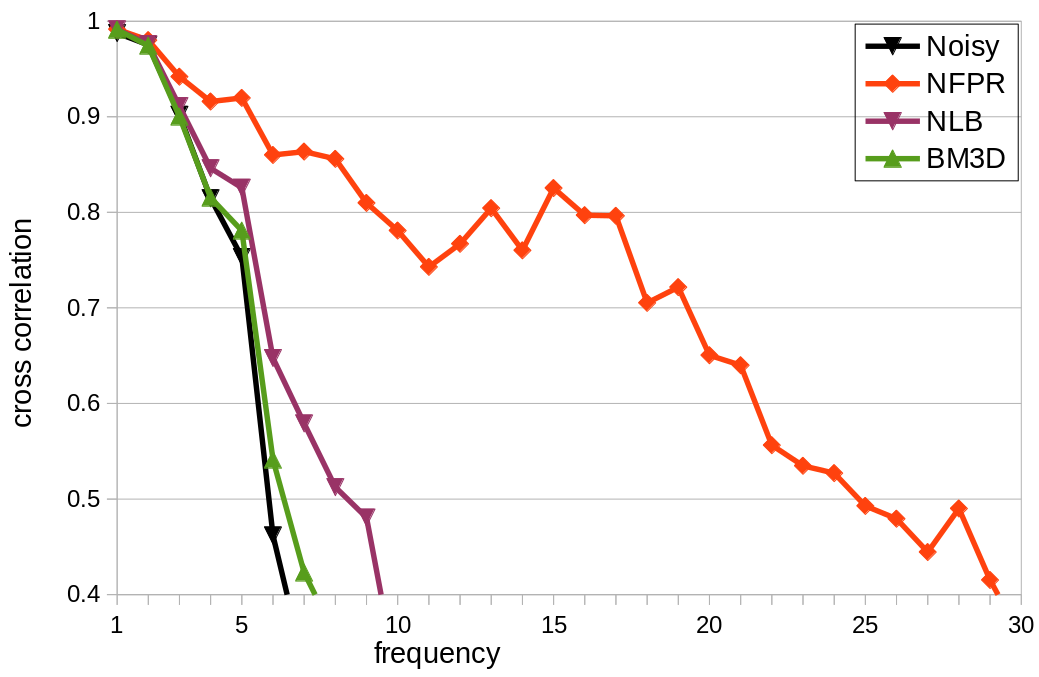} \hspace{1em}	

\caption{\textbf{Left:} 
Zoom into ribosome image of a yeast cell, 
with original size 256 $\times$ 256.  
\textbf{Right:} Zoom into the  
[0.4-1.0] correlation range of the corresponding FRC plot 
calculated for 64 frequency levels.
NFPR parameters used:-
$\sigma = 170, \lambda = 2.5, k_\textrm{max} = 35$. 
For NLB and BM3D we have optimised the unknown 
$\sigma_{\textrm{noise}}$ with respect to FRC.}
\label{fig:res3}
\end{figure*}
Let us now discuss the details of the above two individual terms:
The functions $g$ and $h$ model the above mentioned 
tonal and neighbourhood similarity assumptions, respectively. 
However, if we look closely at the argument of $g$, we have 
${u_{\sigma j}} - {u_{\sigma i}}$ instead of 
${u_{ j}} - {u_{i}}$ which are the real tonal differences.
This idea of calculating the tonal differences on a denoised image
$\bm{u}_{\sigma}$, for a robust performance, is inspired from 
diffusion-based methods \cite{CLMC92}. We have chosen a 
collaborative version of the non-local means approach \cite{BCM05} 
for this initial denoising process:
\begin{align*}
\label{discretization_1}
u^{k}_{\sigma i} =  
b_i^k \cdot \left( \sum_{\substack{j \in P_i^k} } 
 h\left( d_{ij}^k \right) u_j^k
+ \sum_{\substack{j \in P^{\textrm{add},k}_{i}} }
 h \left( d_{ij}^k \right) u_j^k  \right)
\numberthis.
\end{align*}
The symbols $a_i$ and $b_i$ in \eqref{discretization} and 
\eqref{discretization_1}, respectively, are the normalisation 
constants. The functions $g$ \cite{We97} 
and $h$ in \eqref{discretization} are chosen as 
\begin{gather*}
\label{diffusivity}
g\left( s \right) = 1 - \text{exp}\left( 
\frac{-3.31488}{\left(\frac{s}{\lambda}\right)^8} \right), \\
h(s) = \text{exp}\left(\frac{-s^2}{2\sigma^2}\right).
\end{gather*}
The time step size $\tau$ in \eqref{discretization} is selected such 
that the maximum-minimum principle is not violated.
This means that the dynamic range of the denoised image does not 
exceed that of the initial noisy image. We can achieve this by choosing 
$a_i$ = $b_i/M_i$, where $M_i$ is the sum of number of elements in 
$P_i$ and $P_i^{\textrm{add}}$, and $\tau \le 1$. 
Finally, we iterate NFPR for $k_\textrm{max}$ times.
We initialise the non-linear smoothing with the 
initial noisy image $\bm{f}$ and the patch reordering 
using a Gaussian smoothed version of $\bm{f}$ with standard 
deviation $\sigma_G$.


\section{Experiments and Discussion}
\label{sec:expAndDisc}

In order to test the robustness of our method with respect 
to the noise type, we have performed denoising experiments on both 
real-world test images and electron microscopy data. 
For saving time, we have restricted 
the usage of our patch reordering step to just two iterations.
This has negligible effect on the denoising output.
We have fixed the following parameters: 
$\rho_\textrm{search} = 10$, $\rho_\textrm{sim} = 10$, 
$\sigma_G = 2.5$, $\tau = 0.95$ and the number of elements 
in the reordered set $N = 35$. 
In order to have a correspondence for the parameter $\sigma$ 
between real-world and electron microscopy data, we have 
performed an affine rescaling of the distances $d_{ij}$ 
within the set $P_i$, to [0, 255]. Thus, we just  
optimise the parameters $\sigma$, $\lambda$ 
and $k_\textrm{max}$. 
As already mentioned, our denoising experiments employ
NLB, BM3D, Ram et al. \cite{ram2013image} for comparison purposes  
with available implementations and detailed parameter studies 
in \cite{LBM2013a, Lebrun2012, BW20, ram2013image}.

We first present our results on the real-world images Lena, Bridge, 
House and Peppers\footnote{http://sipi.usc.edu/database/}, which 
have been corrupted with AWGN. We use the mean squared error (MSE) 
for both measuring the quality of the denoised images and 
optimising the parameters. 
Figure \ref{fig:res1} and Table \ref{table1}
show the comparison with NLB and BM3D. 
Visual advantages in terms of less artefacts and more pleasant 
edges are larger than MSE advantages would suggest.
From Table \ref{table2}, we can conclude that our method 
is competetive with the approach of Ram et al. \cite{ram2013image}. 

Experiments on a GPU\footnote{NVIDIA GeForce GTX 970 graphics 
card using C++ and CUDA} show that our method 
takes just 2 and 6.5 seconds for denoising the $256 \times 256$ 
sized House and $512 \times 512$ sized Lena images 
($\sigma_{\textrm{noise}}=100$), respectively.
This implementation could further be improved with pre-computing 
the weighting functions and also through faster implementations of 
patch-similarity computations. 
The available non-parallel CPU\footnote{Intel(R) Core(TM) i7-6700 CPU 
3.4 GHz machine using MATLAB/C} implementation 
of \cite{ram2013image} takes 1128 and 7021 seconds in the 
above scenarios. This is very expensive, even if we take into 
account the technical differences in the implementations. 

We have also considered ribosomal data in  
yeast cells acquired using an electron microscope.
For measuring the quality of the denoised images, 
we use a popular frequency domain measure in electron microscopy 
called Fourier ring correlation (FRC). It computes 
a cross-correlation coefficient between the denoised versions 
of two different images of the same scene at different frequency 
levels \cite{SB1982, P2010}. 
Figure \ref{fig:res3} shows the corresponding results. 
We observe higher correlation coefficients for NFPR in the 
FRC curves. This indicates that it does a better job in 
preserving image structures during the denoising process. 

All the above results can be attributed 
to the previously mentioned modelling advantages of NFPR. 
In contrast to NLB and BM3D, it also benefits by avoiding 
an explicit AWGN noise approach. 
This leads to even better NFPR results for electron microscopy 
data, as such kind of data is generally approximated with a 
more general additive noise model (see Chapter 11 of \cite{Fr2013}).
On the other hand, there are certainly some 
cases when NLB, BM3D and Ram et al. \cite{ram2013image} 
are better than NFPR for real-world images like Lena and 
Bridge which have some amount of texture. 

We observe that an important message from the above results
is in accordance with the conclusion in our 
multi-frame denoising research \cite{BW20}: 
The process of choosing the combination of pixels that undergo 
non-linear smoothing is as important as choosing the type of
non-linearity itself. In BM3D and NLB, we filter  
a group of similar patches altogether. 
In NFPR, we utilise a carefully chosen set 
of pixels and subsequently filter them using a robust procedure. 
This is superior although we do not use any explicit spatial 
context in the filtering process. 
Unlike NLB and BM3D, we apply it only for patch reordering. 
In \cite{BW20}, we observed that a linear temporal filter can 
outperform a non-linear one, but only if the pixels 
that undergo the denoising process are chosen carefully. 


\section{Conclusions and Outlook}
\label{sec:concAndOutlook}

Although most people would agree that artifact avoidance is
desirable in denoising, this goal has hardly been addressed
in practice. It appears that smooth patch reordering is
helpful in that aspect, but its computational workload is
usually very high.
The message of our paper is that this patch reordering can
be fairly simplistic and therefore fast, provided that one
comes up with more sophisticated non-linear filters that
reward pixel and patch similarities in a multiplicative
manner. Moreover, refraining from explicit noise models
can be beneficial in real-world applications that may
deviate from ideal noise assumptions. We believe that these
are fairly general design principles that deserve further
exploration in the future. This is also part of our
ongoing work. 


%
\bibliographystyle{IEEEbib}
\bibliography{myrefs}

\end{document}